\documentclass[manuscript]{aastex} %single column double-spaced document
\newcommand{\myemail}{stiwari@prl.res.in}
\shorttitle{}
\shortauthors{Tiwari, Venkatakrishnan and Gosain}

\begin{document}

\title{Magnetic Non-Potentiality of Solar Active Regions and Peak X-Ray
Flux of the Associated Flares}

\author{Sanjiv Kumar Tiwari,
        P.~Venkatakrishnan and Sanjay Gosain}
\affil{Udaipur Solar Observatory, Physical Research Laboratory,
 Dewali, Bari Road,\\
 Udaipur-313 001, India}
\email{\myemail} %stiwari@prl.res.in}
\email{pvk@prl.res.in}
\email{sgosain@prl.res.in}

\begin{abstract}

Predicting the severity of the solar eruptive phenomena like flares
and Coronal Mass Ejections (CMEs) remains a great challenge despite
concerted efforts for several decades. The advent of high quality
vector magnetograms obtained from Hinode (SOT/SP) has increased the
possibility of meeting this challenge. In particular, the Spatially
Averaged Signed Shear Angle (SASSA) seems to be an unique parameter
to quantify the non-potentiality of the active regions. We
demonstrate the usefulness of SASSA for predicting the flare
severity. For this purpose we present case studies of the evolution
of magnetic non-potentiality using 115 vector magnetograms of four
active regions namely ARs NOAA 10930, 10960, 10961 and 10963 during
December 08$-$15, 2006, June 03$-$10, 2007, June 28$-$July 5, 2007
and July 10$-$17, 2007 respectively. The NOAA ARs 10930 and 10960
were very active and produced X and M class flares respectively,
along with many smaller X-ray flares. On the other hand, the NOAA
ARs 10961 and 10963 were relatively less active and produced only
very small (mostly A and B-class) flares. For this study we have
used a large number of high resolution vector magnetograms obtained
from Hinode (SOT/SP). The analysis shows that the peak X-ray flux of
the most intense solar flare emanating from the active regions
depends on the magnitude of the SASSA at the time of the flare. This
finding of the existence of a lower limit of SASSA for a given class
of X-ray flare will be very useful for space weather forecasting. We
have also studied another non-potentiality parameter called mean
weighted shear angle (MWSA) of the vector magnetograms along with
SASSA. We find that the MWSA does not show such distinction as the
SASSA for upper limits of GOES X-Ray flux of solar flares, however
both the quantities show similar trends during the evolution of all
active regions studied.
\end{abstract}

\keywords{Sun: flares, Sun: magnetic fields, Sun: sunspots}

\section{Introduction}

Many magnetic parameters e.g., twist, shear, energy etc. computed
for complex active regions have been examined with a view to predict
the severity of solar eruptive phenomenon like flare and Coronal
Mass Ejections (CMEs). Although such studies have been carried out
for several decades, the progress remains slow. Here we study two of
these parameters i.e., magnetic non-potentiality inferred from the
spatially averaged signed shear angle (SASSA) and mean weighted
shear angle (MWSA) of sunspots.

Magnetic shear at polarity inversion lines were studied earlier to
look for flare related changes
\citep[e.g.,][]{hagy84,hagy90,amba93,hagy99b}. The magnetic energy
change following few flares has been estimated by
\cite{schr08,jing10}. \cite{jing10} found that the magnitudes of
free magnetic energy were different for the flare-active and the
flare-quiet regions, but the temporal variation of free magnetic
energy did not show any clear and consistent pre-flare pattern. A
study of the evolution of global alpha ($\alpha_g$) for one highly
eruptive and one quiet active region has been performed by
\cite{tiw09e}. But no correlation between $\alpha_g$ and the GOES
X-Ray flux was found. \cite{sudo05} found changes in the line of
sight magnetic field associated with flares. The unsigned flux of
NOAA AR 10930 has been studied recently by \cite{abra08}. Figure 2
of \cite{abra08} shows that the unsigned net flux of NOAA AR 10930
does not show any relationship with the peak GOES X-ray flux.

Despite many such attempts, the severity of flares has not been
successfully predicted.
In this paper, we make an effort to determine an upper limit of the
peak X-ray flux of solar flares emanating from the active
regions as a function of the magnitude of the SASSA and the MWSA.
These parameters, {\it viz.,} SASSA and MWSA give quantitative
measure of the non-potentiality present in an active region at the
observed height. More details about the SASSA and the MWSA are given
in Section 2.

Many researchers have used the force-free parameter $\alpha$ as a measure of the
magnetic twist of the sunspots. Many forms of global $\alpha$ have been proposed
and studied such as:
$\alpha_{best}$ \citep{pcm95}, $\alpha_{av}$ \citep{hagi04},
$\alpha_g$ \citep{tiw09a,tiw09b,tiw09e} etc. The force-free parameter $\alpha$
actually gives twice the degree of twist per unit axial length along the axis of
the flux rope (see Appendix A of \cite{tiw09a}).
Thus, $\alpha$ provides the gradient of twist at a certain observational height
and not the actual twist of an active region.
The sign of the global $\alpha$ ($\alpha_g$) and the SASSA are found similar
but the magnitudes are not correlated \citep{tiw09b}.
Recently, $\alpha_g$ has been studied for the time series of two
active regions NOAA AR 10930 and 10961 by \cite{tiw09e} as mentioned earlier.
The global alpha does not show any clear indication for predicting
the severity of the X-ray flares. The reason could be
due to the non-validity of the linear force-free assumption for
such complex ARs.
The local and global $\alpha$ values of several active regions
were studied by \cite{hahn05} and \cite{nand03,nand08}.
They found that the global $\alpha$ ($\alpha_{best}$)
was not important for the flare activity. However, they noticed a
decrease in the variance of spatial $\alpha$ distribution after the
flare.

To explore the utility of SASSA as a predictor of flare severity, we
have studied the evolution of SASSA in a time series of vector
magnetograms of two highly flare productive sunspots i.e., NOAA ARs
10930 and 10960 and also two less flare productive sunspot i.e.,
NOAA ARs 10961 and 10963. The AR 10930 has been the most active
sunspot observed by Hinode (SOT/SP). Three major X-class flares
i.e., X6.5, X3.4 and X1.5 were observed by Hinode (SOT/SP) on 06, 13
and 14 December, 2006 respectively. Many C and B-class flares were
also associated with the same sunspot. Similarly NOAA AR 10960 was
also highly flare productive and produced four M-class flares. On
the other hand NOAA ARs 10961 and 10963 were relatively less flare
productive.

The main purpose of this study is to find the lower limit of the
non potentiality parameters, if any, for a given class of X-Ray flare.

In the following section (Section 2), we briefly describe the magnetic
parameters SASSA and MWSA. In Section 3, we give details of the
data sets used. Section 4 illustrates the analysis and results
obtained. Finally, in Section 5 we discuss the results and
present our conclusions.

\section{The Magnetic Non-Potential Parameters Used in This Study}

\subsection{Spatially Averaged Signed Shear Angle (SASSA)}

Signed shear angle (SSA) represents the deviation of observed
transverse vectors from the potential transverse vectors with positive
or negative sign. It has the similar sign as the photospheric
chirality of the sunspots \citep{tiw09e,tiw09b}.
The SSA is computed from the following formula:
\begin{equation}\label{}
SSA = \tan^{-1} (\frac{B_{yo} B_{xp} - B_{yp} B_{xo}}{B_{xo} B_{xp} + B_{yo} B_{yp}})
\end{equation}
where $B_{xo}, B_{yo}$ and $B_{xp}, B_{yp}$ are observed and potential transverse
components of sunspot magnetic fields respectively.

A spatially averaged value of SSA (SASSA) is taken to quantify the global
non-potentiality of the whole sunspot. This parameter gives the
non-potentiality of a sunspot irrespective of the force-free
nature \citep{tiw09b} and shape of the sunspot \citep{venk09}.
Thus, SASSA is a very important magnetic parameter to quantify the
non-potentiality of any active region.

Apart from measuring the non-potentiality, SASSA also retains the
sign of chirality of the active region magnetic field, unlike other
shear parameters such as MWSA. This property of sign seems to be
crucial in obtaining the proper measure of global non-potentiality,
as will be explained later in the paper.

\subsection{Mean Weighted Shear Angle (MWSA)}

The mean weighted shear angle was introduced by \cite{wang92} to
quantitatively study the changes in magnetic structure and the build-up of the
magnetic shear. The mean weighted shear angle is given as
\begin{equation}\label{}
MWSA = \frac{\sum |B_t| ~\theta} {\sum |B_t|}
\end{equation}
where $B_t$ is the measured transverse field strength and $\theta$ is the
difference between the observed and potential azimuths. The potential fields
have been computed by taking the longitudinal field as boundary.
The method used in computing the potential field is as per \cite{saku89}.

The reason for calculating weighted mean instead of a simple average
of shear angle is that the MWSA filters the weak field area. The
stronger fields play more important role in determining the field
structure and can be measured more accurately. We should note here
that the MWSA will weight more on the high transverse field regions
like penumbral fields, a fact that will be shown later in the paper
to explain the relatively lower success of MWSA as a flare intensity
predictor.

\section{Data Sets Used}

We have used the series of vector magnetograms of two
eruptive ARs NOAA 10930 and 10960 and two less-eruptive ARs NOAA
10961 and 10963 obtained from the Solar Optical Telescope/Spectro-polarimeter
(SOT/SP: \cite {tsun08,suem08,ichi08,shim08}) onboard Hinode \citep{kosu07}.

The Hinode (SOT/SP) data have been calibrated by the standard
``SP\_PREP'' routine developed by B. Lites and available in
the Solar-Soft package. The prepared polarization spectra have
been inverted to obtain vector magnetic field components using an
Unno-Rachkowsky \citep{unno56,rach67}
inversion under the assumption of Milne-Eddington (ME)
atmosphere \citep{lando82,skum87}. We use the ``STOKESFIT"
inversion code which is available in the Solar-Soft package and was
developed by T. R. Metcalf. The latest version of the
inversion code is used which returns the true field strengths along
with the filling factor.

There is an inherent 180$^{\circ}$ ambiguity in the azimuth
determination due to the insensitivity of the Zeeman effect to the
sense of orientation of the transverse magnetic fields. Numerous
techniques have been developed and applied to resolve this problem
\citep[for details see][]{metc06,leka09}, but a complete resolution
is not expected from the physics of the Zeeman effect. The chirality
of chromospheric and coronal structures can be used as guides to
complement the other methods. The 180$^{\circ}$ azimuthal ambiguity
in our data sets have been removed by using the acute angle method
\citep{harv69,saku85,cupe92}. This method of ambiguity resolution
works very well for magnetic shear angles that are less than 90
degrees. Less than one percent pixels of any vector magnetogram
studied has shear $\sim 90$ degrees. Therefore, we expect that the
acute angle method works well in all our cases. Most of the data
sets used have a spatial sampling of $\sim0.3$ arcsec/pixel. A few
data sets are observed in ``Normal Mode" of SOT with a spatial
sampling of $\sim0.16$ arcsec/pixel.

The noise in the data has been minimized in the similar way as was
done in \cite{tiw09e,tiw09b} and \cite{venk09,venk10}.
The method is as follows:
the pixels having transverse $(B_t)$ and longitudinal magnetic
field $(B_z)$ values greater than a certain level are only
analyzed. To decide this critical threshold, a quiet Sun region
is selected for each active region and the 1$\sigma$ deviation in the
three vector field components $B_x$, $B_y$ and $B_z$ are
evaluated separately. The resultant 1$\sigma$ deviation in $B_x$ and $B_y$
is then taken as the noise level for transverse field
components. Only those pixels with longitudinal and transverse
fields simultaneously greater than twice the
above mentioned noise levels are analyzed.

To minimize the projection effects, the magnetograms are transformed
to the heliographic coordinates wherever they are more
than 10 degrees away from the disk-center \citep{vhh88,venk89}.

The information about different classes of X-ray flares are
collected from the web sites \url{http://www.solarmonitor.org/index.php} and \url{http://www.spaceweather.com/}.
We have also used the GOES X-Ray data.

\section{Data Analysis and Results}

Figure 1 shows GOES 12 X-Ray plots (in the wavelength range 1.0$-$8.0 \AA)
of different X-Ray flares observed during the disk passage of NOAA
ARs 10930, 10960, 10961 and 10963 respectively.
We see that the GOES X-ray peaks show high activity with two X-class
flares and several C-class flares during 08$-$15 December, 2006 in
NOAA AR 10930 (Figure 1(a)).
High activity is also seen with 4 M-class and several C-class flares
during 03$-$10 June, 2007 through the disk passage of NOAA AR 10960 (Figure 1(b)).
The GOES X-ray light curves show lesser activity with very small B and A class flares
during June 28$-$July 05, 2007 in NOAA AR 10961 (Figure 1(c)).
Figure 1(d) also presents a weak activity with some C-class flares in the beginning
but mostly B and A-class flares in later evolution stages of NOAA AR 10963
during 10$-$18 July, 2007.

A large number of vector magnetograms of eruptive NOAA ARs 10930
(36) and 10960 (21) have been analyzed. Also, 28 and 30 vector
magnetograms of lesser eruptive NOAA ARs 10961 and 10963
respectively, are analyzed to study the evolution of these sunspot
magnetic fields. One longitudinal image with the transverse vectors
of each active region has been shown in the Figure 2. The red and
blue contours represent the positive and negative fields of
$\pm1000, \pm1500, \pm3000$ Gauss.

\subsection{Temporal Evolution of SASSA and MWSA}

Figures 3 and 4 show the plots of SASSA and MWSA in highly active regions
NOAA ARs 10930 and 10960 and relatively quiet NOAA ARs 10961 and 10963 respectively.
The diamonds and stars represent the SASSA and MWSA respectively.
The dotted/dashed vertical lines show the time and class of flares associated
with these active regions.
The lines with red colored big dashes in Figure 3(a) represent the timings of
X-class flares. The orange colored dash-dotted lines in Figure 3(b) show the
timings of M-class flares. The lines with purple colored small dashes in all
figures represent the timings of C-class flares. The blue colored dotted lines
in Figures 3(a) and 3(b) represent the timings of B-class flares.
The black colored dotted lines in Figures 4(a) and 4(b) represent the timings
of A-class flares whereas blue colored dashed lines in these figures represent
the timings of B-class flares.

\subsubsection{Evolution of SASSA and MWSA in NOAA AR 10930}

From the plots of SASSA in Figure 3(a), it can clearly be noticed
that any X-class flare happened only when the SASSA was greater than
$8$ degrees. If the SASSA is greater than $4$ degree, then C-class
flares also occurred. If the SASSA is less than $4$ degrees, only
B-class flares happened. It should be noted that the sign of SASSA
with magnitudes is given in the Figures only to express the sense of
chirality of non-potentiality. Similarly, the MWSA was greater than
26 and 22 degrees for X and C-class flares respectively. B-class
flares  occurred if the MWSA was greater than 21 degrees.

We can notice that both the non-potentiality parameters show similar trend
during the evolution period of the active region. However, the threshold
values of MWSA for C and B-class flares are very close.

\subsubsection{Evolution of SASSA and MWSA in NOAA AR 10960}

Figure 3(b) shows that the M-class flares took place when SASSA exceeded
6 degrees. The M-class flares did not occur whenever SASSA was less than 6 degrees.
One interesting behavior during the evolution of this active region can be noticed
is the chirality inversion. The SASSA in the beginning is highly negative and three
M-class flares took place. The SASSA started decreasing allowing only 3 C-class flares
and became positive by building up non-potentiality again. As SASSA became more than 4
degrees, C-class flares again started taking place. After 7 C-class flares, one M-class
flare happened when SASSA exceeded 6 degrees. Even in this peculiar evolution
of the active region, the SASSA maintained its upper limits for different classes of
X-ray flares as observed in the NOAA AR 10930. Thus we see a symmetrical behavior of
the threshold values, independent of the sign of SASSA.

The MWSA shows an almost similar behavior as the SASSA. MWSA was high in the beginning,
starts decreasing after some flares and again builds up showing a similar trend as SASSA.
It can be noticed however that for C-class flares, MWSA goes very small up to 11 degrees
and could not maintain its upper limit as decided in the case of NOAA AR 10930. The
MWSA values observed for M-class flares are seen to be more than 22 degrees which mixes
up with that of the C-class flares observed in NOAA AR 10930.
Thus, the threshold values of MWSA seem to be specific to the active region and thus
seem to be of dubious utility for flare intensity prediction.

\subsubsection{Evolution of SASSA and MWSA in NOAA AR 10961}

The NOAA AR 10961 is the quietest active region out of the four studied.
Only B and A-class flares occurred during the disk passage of this active
region as can be seen in Figure 4(a).
From Figure 4(a), one can see that the B-class flares occurred in NOAA AR 10961
when the SASSA was greater than $\sim 2.5$ degrees.
Only small A-class flares happened when SASSA was less than $2$ degrees.

The MWSA was greater than 16 and 14 degrees respectively for any B and
A-class flare to occur. The MWSA shows a very similar trend as the SASSA
but the threshold values again seemed to be specific to the active regions.

\subsubsection{Evolution of SASSA and MWSA in NOAA AR 10963}

The NOAA AR 10963 is relatively quiet active region than the first two cases.
It produced several C-class flares in the beginning of the observations
and later became very quiet triggering only B and A-class flares as shown
in Figure 4(b). From the Figure 4(b), we note that the SASSA was high on
10th July when several consecutive C-class flares took place. No C-class flare
took place when SASSA became less than 4 degrees complementing the behavior
seen in cases of Figures 3(a) and 3(b). No B-class flare occurred when SASSA
was less than 2.5 degrees as observed in the Figure 4(a). Thus SASSA
maintains its threshold values even for very small flares.

The MWSA was observed greater than 20 degrees for C-class flares.
For every B and A-class flare, MWSA was greater than 18 and 16
degrees respectively. The MWSA again shows a similar evolutionary
behavior as that of the SASSA. The threshold values, however, were
specific to the active region.

Thus, we conclude that the SASSA can be used as a reliable predictor
of maximum possible flux of X-ray flares.
The data for studying the evolution of SASSA in more active regions
are not available for the time being. However, the inspection of the
different flares which occurred in all the sunspots listed in the
Table 1 of \cite{tiw09b} show a similar trend.
From the Table 1 of \cite{tiw09b}, we also confirm a threshold value
of $\sim$6 degrees for M-class flares as observed in our Figure 3(b).
The peak flux of M-class flares associated with NOAA AR 10808 and
NOAA AR 09591 are shown with their SASSA and MWSA values respectively
by diamond symbols in Figures 5(a) and 5(b).

\subsection{Statistical relation between the peak X-Ray flux and the
non-potentiality parameters}

Figures 5(a) and 5(b) represent scatter plots between the peak GOES X-ray
flux and interpolated SASSA and MWSA values for that time, respectively.
The cubic spline interpolation of the sample of the SASSA and the MWSA values
has been done to get the SASSA and MWSA exactly at the time of peak flux of
the X-ray flare.

\subsubsection{Statistical relation between the peak X-Ray flux and the SASSA}
From the Figure 5(a), it can be noted that there is a good
relationship between the minimum value of the magnitude of SASSA and
the observed value of the peak GOES X-ray flux. Thus, a lower limit
of SASSA can be assigned for each class of X-ray flare.

\subsubsection{Statistical relation between the peak X-Ray flux and the MWSA}
Figure 5(b) does not show as good correlation as SASSA. The MWSA is
small even for higher classes of flares. One possible reason for
such behavior of MWSA is explained in Section 5.

\section{Discussion and Conclusions}

From the Figures 3 and 4, it is very clear that an upper limit of
peak X-ray flux for a given value of SASSA can be given for
different classes of X-ray flares. Thus, we conclude that the SASSA,
apart from its helicity sign related studies, can also be used to
predict the severity of the solar flares. However to establish these
lower limits of SASSA for different classes of X-ray flares, we need
more cases to study. The SASSA already gives a good indication of
its utility from the present four case studies using 115 vector
magnetograms from Hinode (SOT/SP). Once the vector magnetograms are
routinely available with higher cadence, the lower limit of SASSA
for each class of X-ray flare can be established by calculating the
SASSA in a series of vector magnetograms. This will provide the
inputs to space weather models.

The other non-potentiality parameter MWSA studied in this paper does
show a similar trend as that of the SASSA. The magnitudes of MWSA,
however, do not show consistent threshold values as related with the
peak GOES X-ray flux of different classes of solar flares. One
possible reason for this behavior may be explained as follows: The
MWSA weights the strong transverse fields e.g., penumbral fields.
From the recent studies \citep{su09,tiw09b,tiw09e,venk09,venk10} it
is clear that the penumbral field contains complicated structures
with opposite signs of vertical current and vertical component of
the magnetic tension forces. Although the amplitudes of the magnetic
parameters are found high in the penumbra, they do not contribute to
their global values because they contain opposite signs, which
cancel out in the averaging process \citep{tiw09e,tiw09b}. On the
other hand, the MWSA adds those high values of shear and produces a
pedestal that might mask any the relation between the more relevant
global non-potentiality and the peak X-ray flux. Whereas the SASSA
perhaps gives more relevant value of the shear after cancelation of
the penumbral contribution.

The other parameters such as unsigned net flux and global alpha
($\alpha_g$) do not show good correlation with the GOES X-Ray flux
as discussed in the Introduction (Section 1). However, the evolution
of SASSA in 10930 during 09$-$14 December 2006 shows a good
correlation with the free magnetic energy as seen from the Figure 6
of \cite{jing10}. In particular, SASSA remains about 5 degrees
during 09$-$11 December 2006 and increases up to about 12 degrees by
the end of 13th December 2006. The free magnetic energy (as computed
by \cite{jing10}) follows the similar trend with the values of about
40$\times 10^{31}$ erg during 09$-$11 December 2006 which increases
up to 90$\times 10^{31}$ erg by 14th December 2006. Thus SASSA seems
a good indicator of free magnetic energy as well.

We now proceed to give a plausible reason for the success of SASSA
as a predictor of maximum flare intensity. Basically we need two
ingredients for a flare. We need available magnetic energy
(non-potential field energy) and we also need a flare trigger which
is usually magnetic reconnection driven by the flux emergence. A
flare cannot occur in the absence of either ingredient. However, in
the present study, we find that a single global property of the
spot, viz., SASSA decides the class of flare, if one were to occur.
The class of flare, as classified by soft X-ray emission, indicates
the maximum amount of X-ray emission and therefore maximum mass of
the emitting plasma. Our definition of SASSA is basically an
indication of the amount of non-potentiality in the field
configuration. It has already been established that higher
non-potentiality is related to lower magnetic tension
\citep{venk90,venk93}. Further, lower magnetic tension implies
larger scale heights of the magnetic pressure for a force-free
field. A larger scale height of magnetic pressure means a more
extended field, and thus a larger volume of emitting gas filling
these extended fields by chromospheric evaporation during a flare. A
larger volume of emitted plasma results in higher X-ray emission for
typical densities of the evaporated plasma. This explanation can be
tested by detailed modeling or by observing flares at the solar limb
for sunspots with known amount of SASSA or on the disk using
stereoscopic observations. If this explanation is indeed borne out
by such studies, then it will confirm the importance of SASSA for
the dynamical equilibrium of sunspots.

The peak X-ray emission of flares was also found to be correlated
with the initial speed of the associated CME \citep{ravi04}. The
initial speed of CMEs was in turn found to be related to the
severity of the geomagnetic storm \citep{sriv02}. Thus, SASSA might
also turn out to be a good parameter for predicting the severity of
geomagnetic storms. This will be investigated in future.

Other parameters such as, free magnetic energy and tension forces
can also be studied to examine the pre-flare equilibrium
configuration of a flaring active region. All these quantities, once
studied together in a large number of cases, will certainly provide
a better prediction of flare severity. Monitoring the evolution of
SASSA and other magnetic parameters will require high-cadence vector
magnetograms which can be obtained from filter based instruments
like Solar Vector Magnetograph (SVM:
\cite{gosain04,gosain06,gosain08}), Multi Application Solar
Telescope (MAST: \cite{venk06}) from the ground and Helioseismic and
Magnetic Imager (HMI: \cite{sche02}) aboard the SDO, from the space.

\acknowledgments {\bf Acknowledgements}\\
We thank the referee for very useful comments to improve the
manuscript. The authors acknowledge IFCPAR (Indo French Centre for
Promotion of Advanced Research) for financial support. We
acknowledge SWPC for providing GOES X-ray data. Hinode is a Japanese
mission developed and launched by ISAS/JAXA, collaborating with NAOJ
as a domestic partner, NASA and STFC (UK) as international partners.
Scientific operation of the Hinode mission is conducted by the
Hinode science team organized at ISAS/JAXA. This team mainly
consists of scientists from institutes in the partner countries.
Support for the post-launch operation is provided by JAXA and NAOJ
(Japan), STFC (U.K.), NASA (U.S.A.), ESA, and NSC (Norway).

%\bibliographystyle{apj}
%\bibliography{stiwari}

\begin{figure}
\epsscale{.68}
\plotone{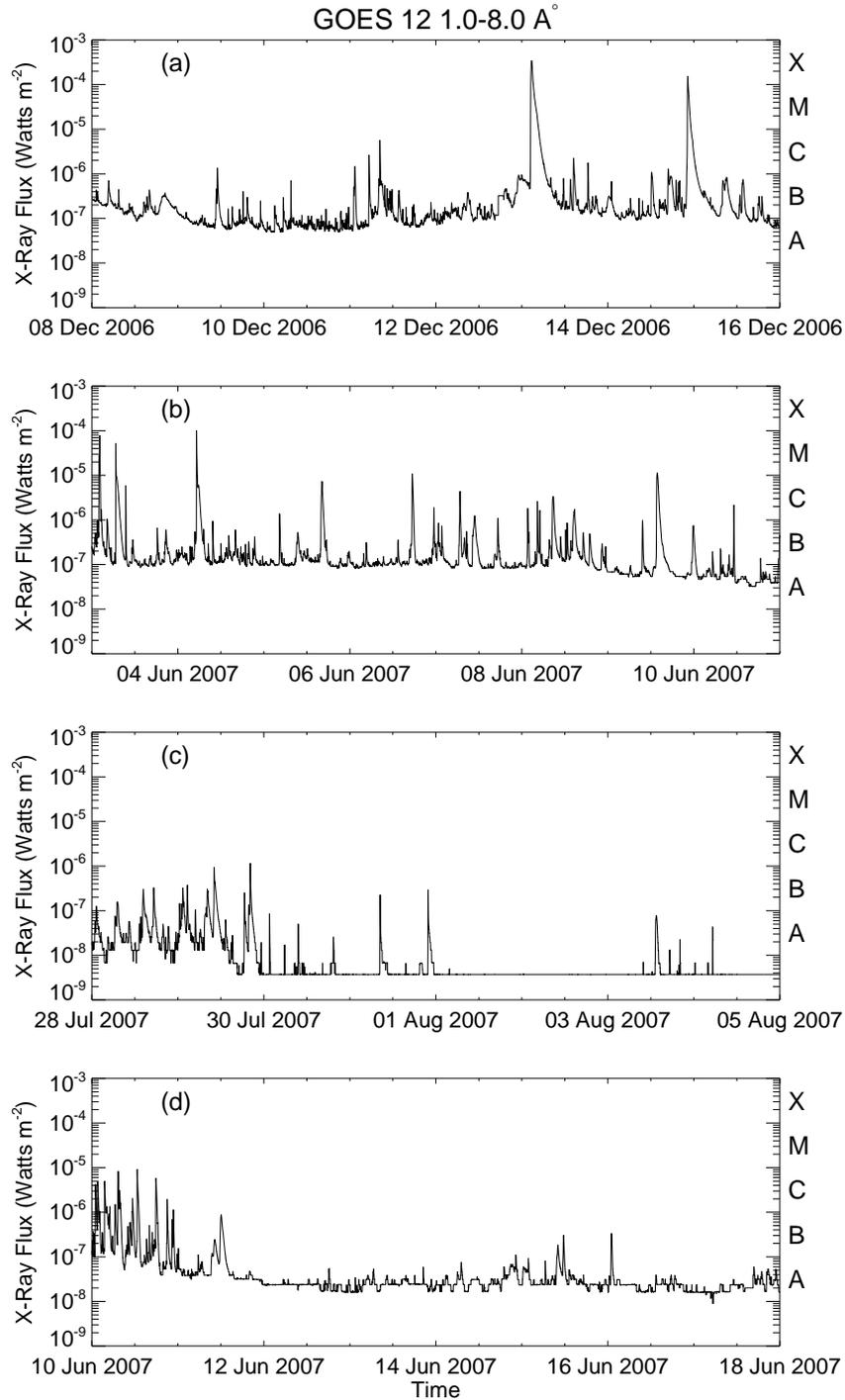}
\caption{Time variation plots of the X-ray flux from GOES 12 satellite in
the wavelength range 1-8 \AA. (a): The data ranges from 09 December 2006 to
15 December 2006 when highly active NOAA AR 10930 was present on the solar disk.
Two X-class and many C \& B-class flares occurred. (b): Four M-class flares and
several C and B-class flares were observed in NOAA AR 10960.
(c) and (d): Both panels show lesser activity with mostly B and A-class flares
in NOAA ARs 10961 and 10963 respectively.}
\end{figure}

\begin{figure}
\epsscale{1.6}
\plottwo{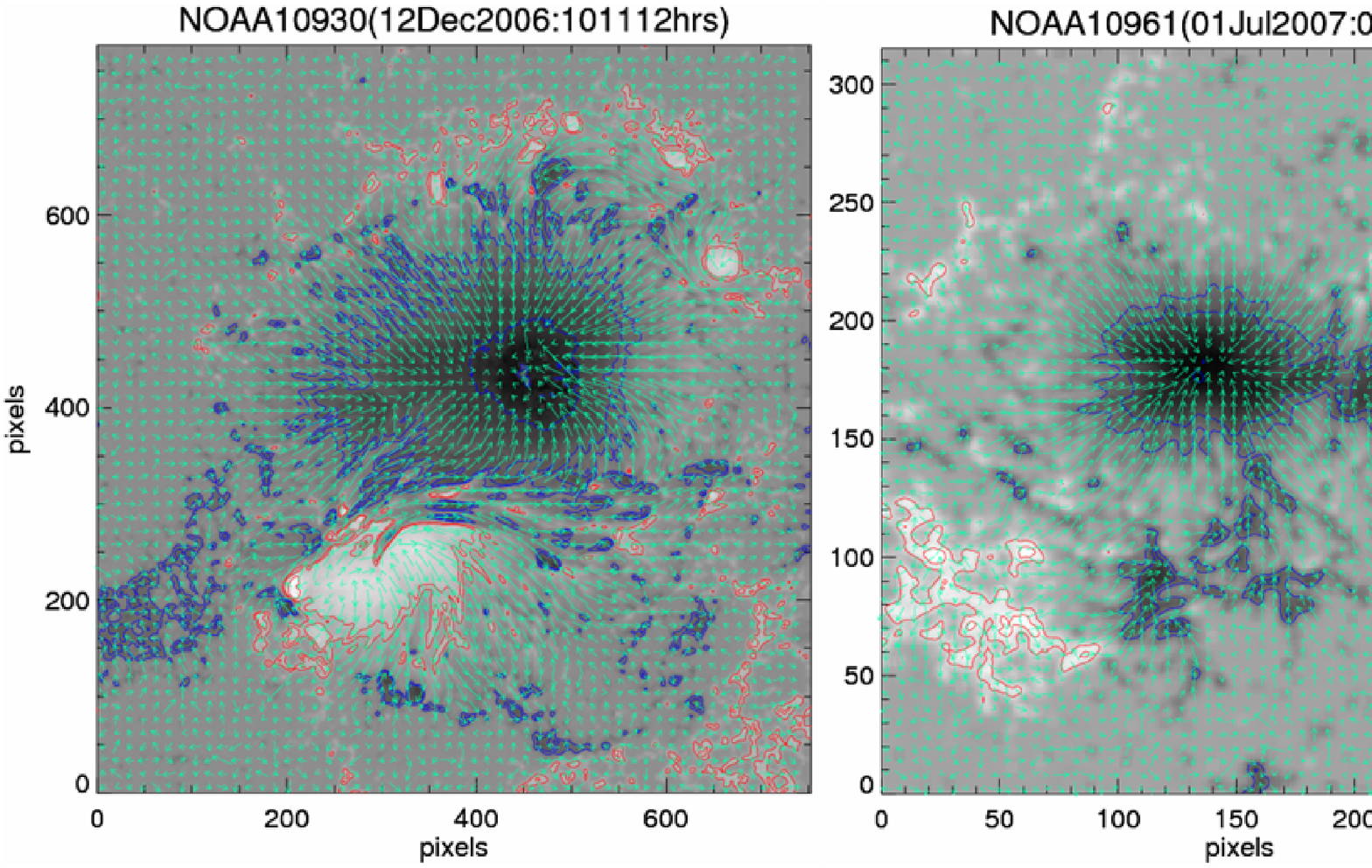}{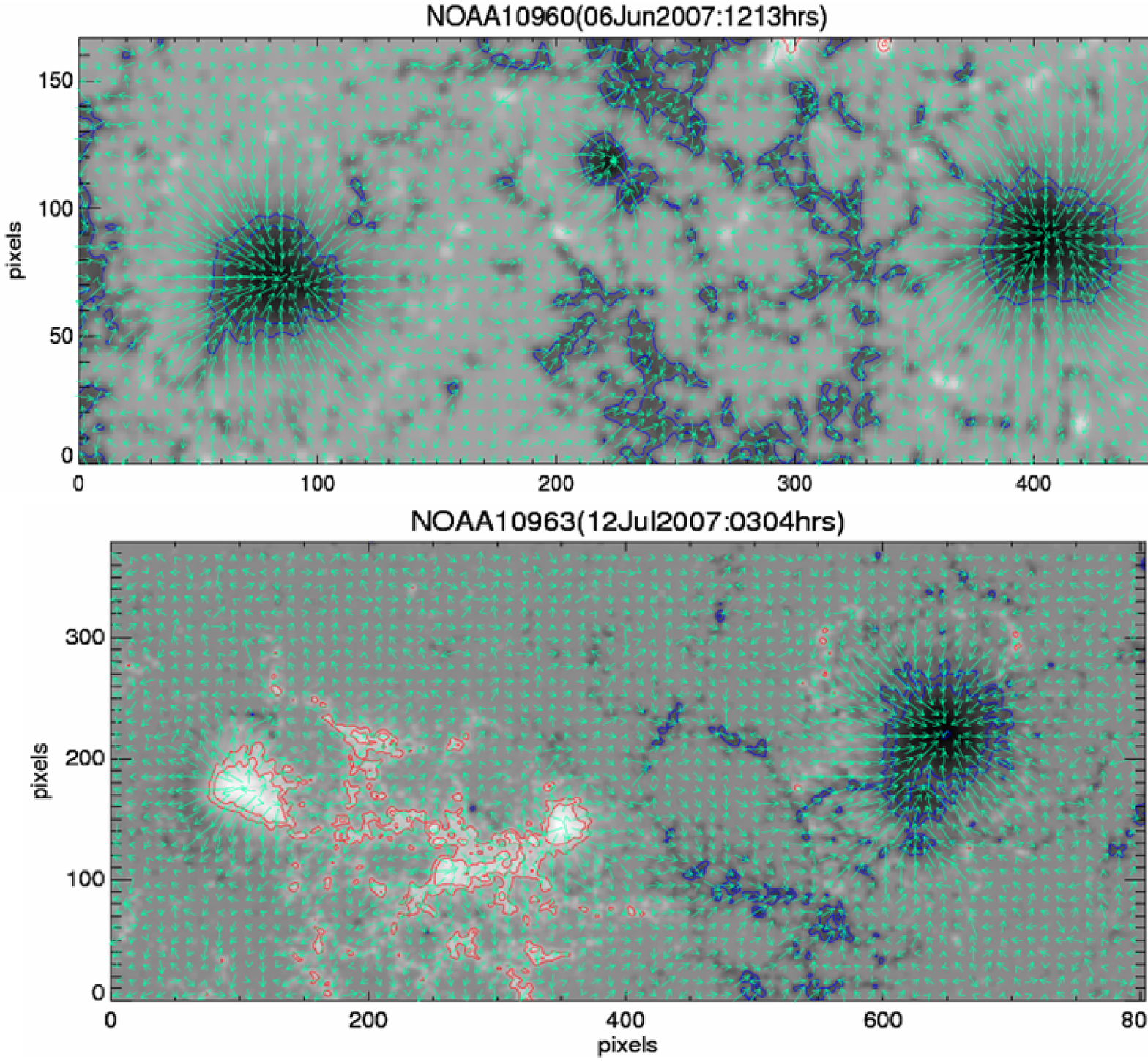}
\caption{Examples of one longitudinal image with the transverse
vectors of each active region studied are shown.
The red and blue contours represent the positive and negative fields
of $\pm1000, \pm1500, \pm3000$ Gauss.}
\end{figure}

\begin{figure}
\epsscale{1.9}
\plottwo{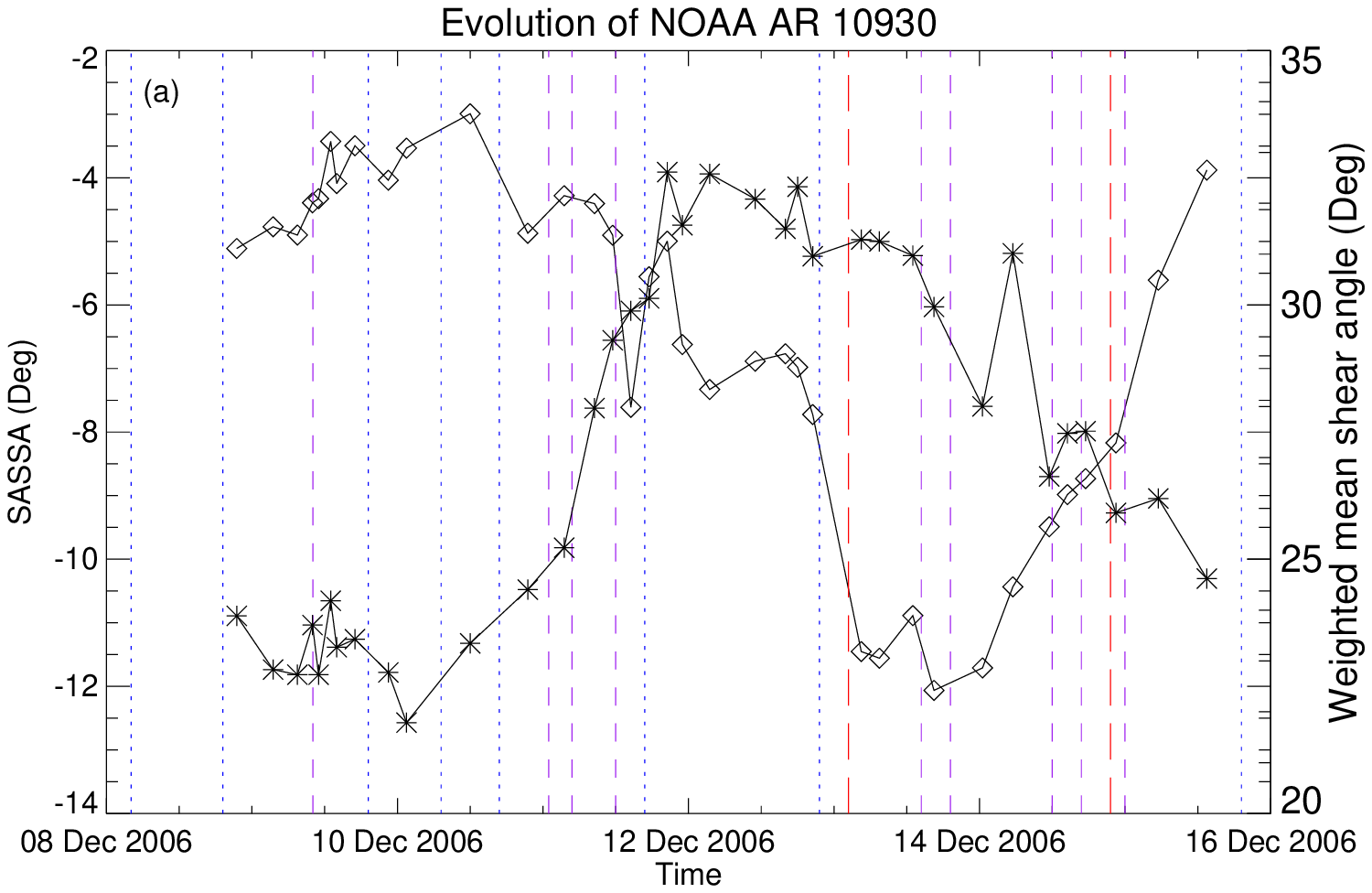}{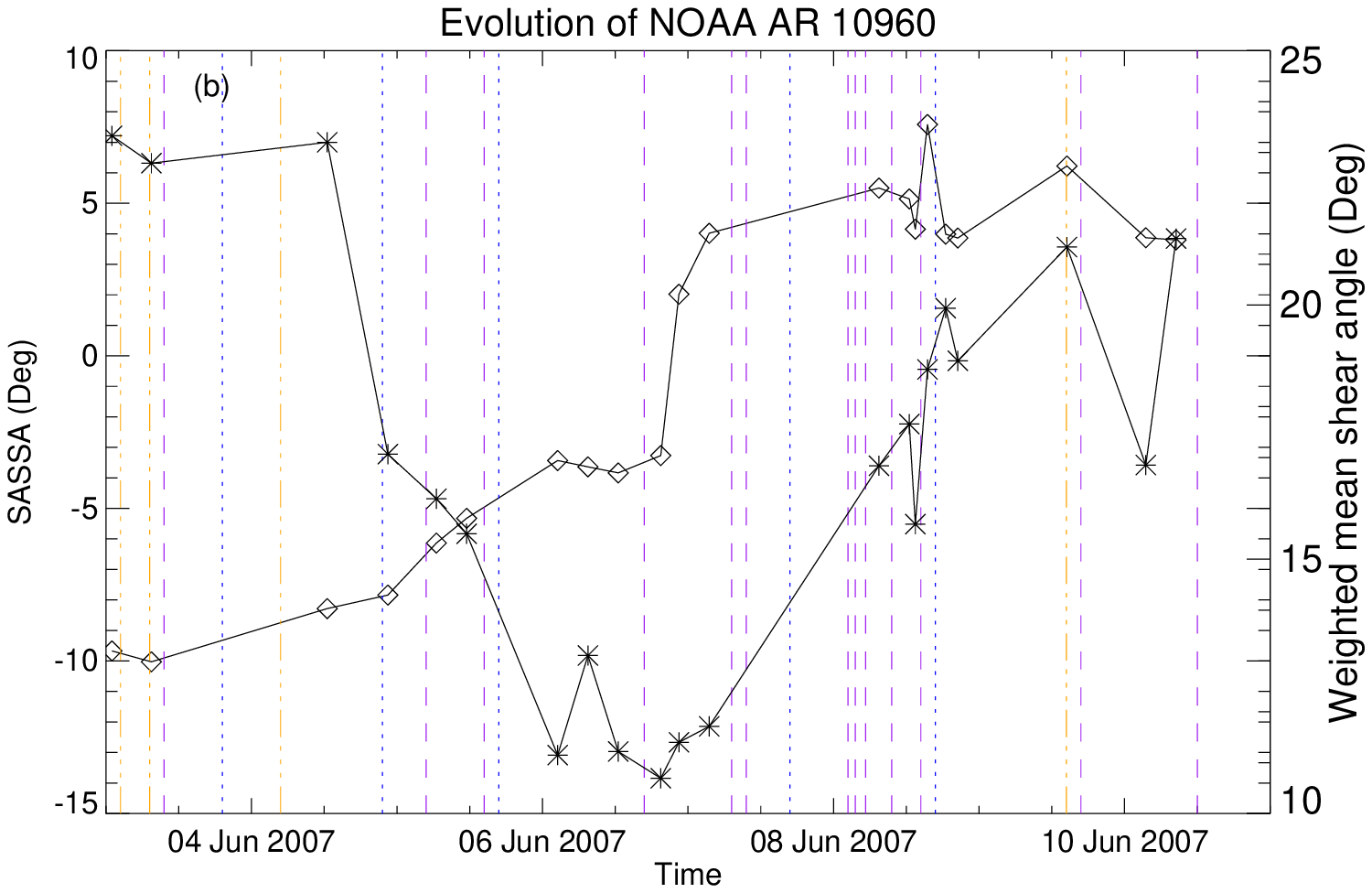}
\caption{Evolution of SASSA and MWSA in NOAA ARs 10930 and 10960 is shown.
The boxes and stars represent the SASSA and MWSA respectively.
(a): The red big dashed vertical lines represent timings of two X-class flares.
Purple colored smaller dashed lines show the timings of C-class flares and
blue dotted lines represent the timings of the B-class flares.
(b): The orange colored dash-dotted lines show the timings of the M-class flares.
The dashed purple and dotted blue lines again show the timings of C and B-class
flares as above.
}
\end{figure}

\begin{figure}
\epsscale{1.9}
\plottwo{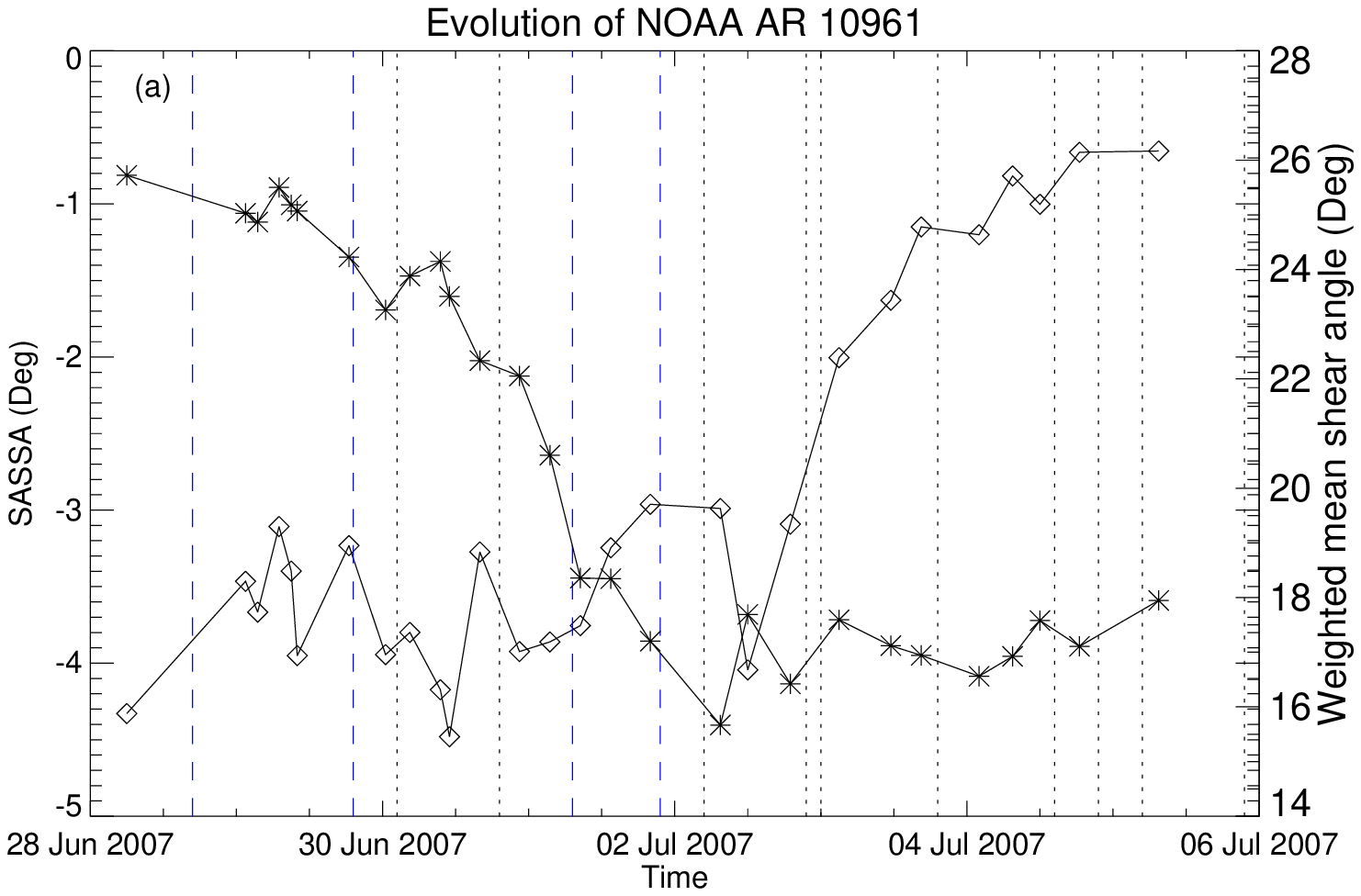}{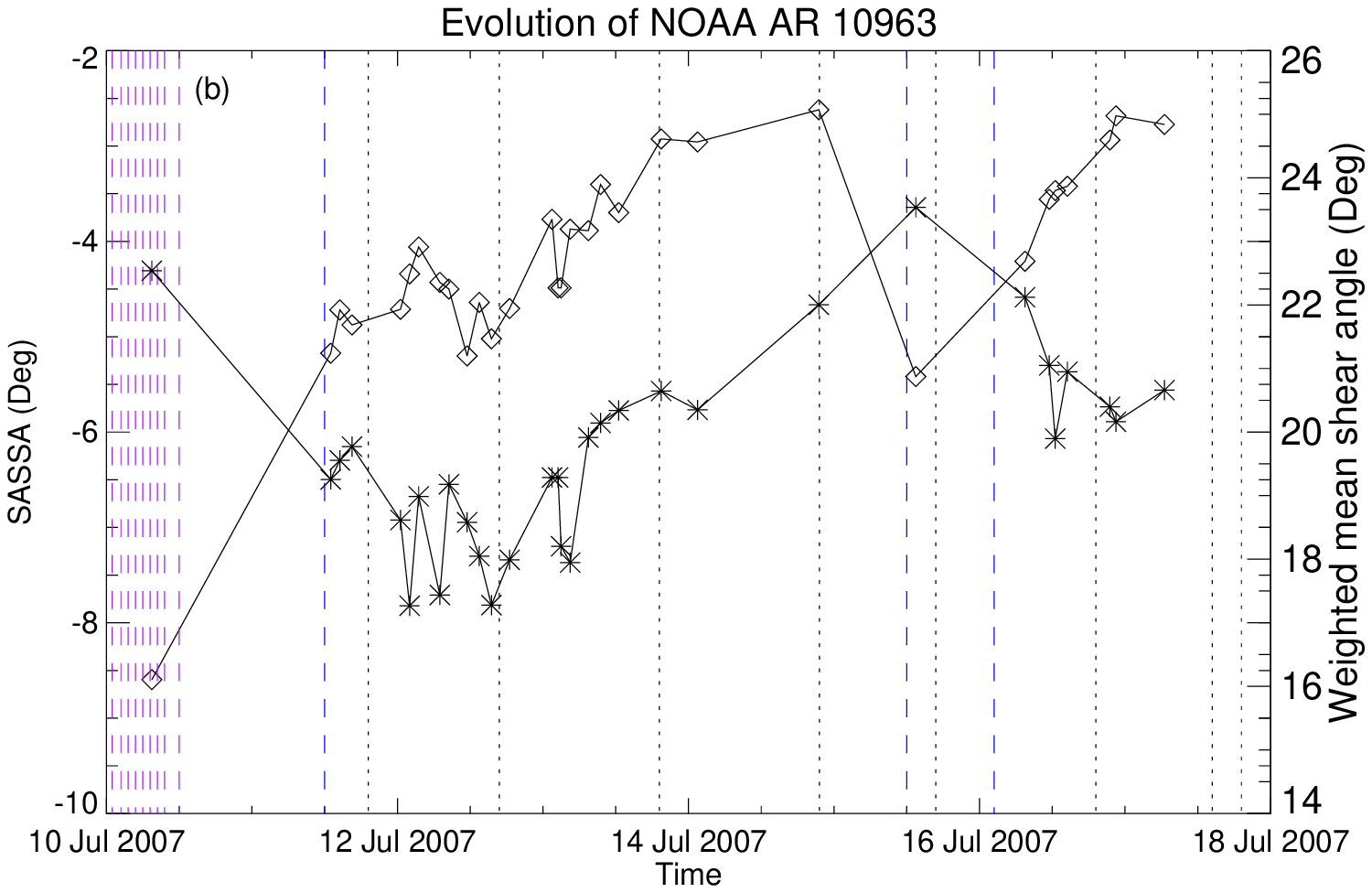}
\caption{Evolution of SASSA in NOAA AR 10961 and NOAA AR 10963 shown. Both are
relatively quiet regions. The boxes and stars represent the SASSA and MWSA respectively.
(a): The blue colored dashed vertical lines represent the timings of the B-class
X-ray flares and black dotted lines represent the timings of A-class flares.
(b): The purple colored dashed lines show timings of C-class flares. The blue colored dashes and black
dots represent timings of B and A-class flares respectively.
}
\end{figure}

\begin{figure}
\epsscale{0.75}
\plotone{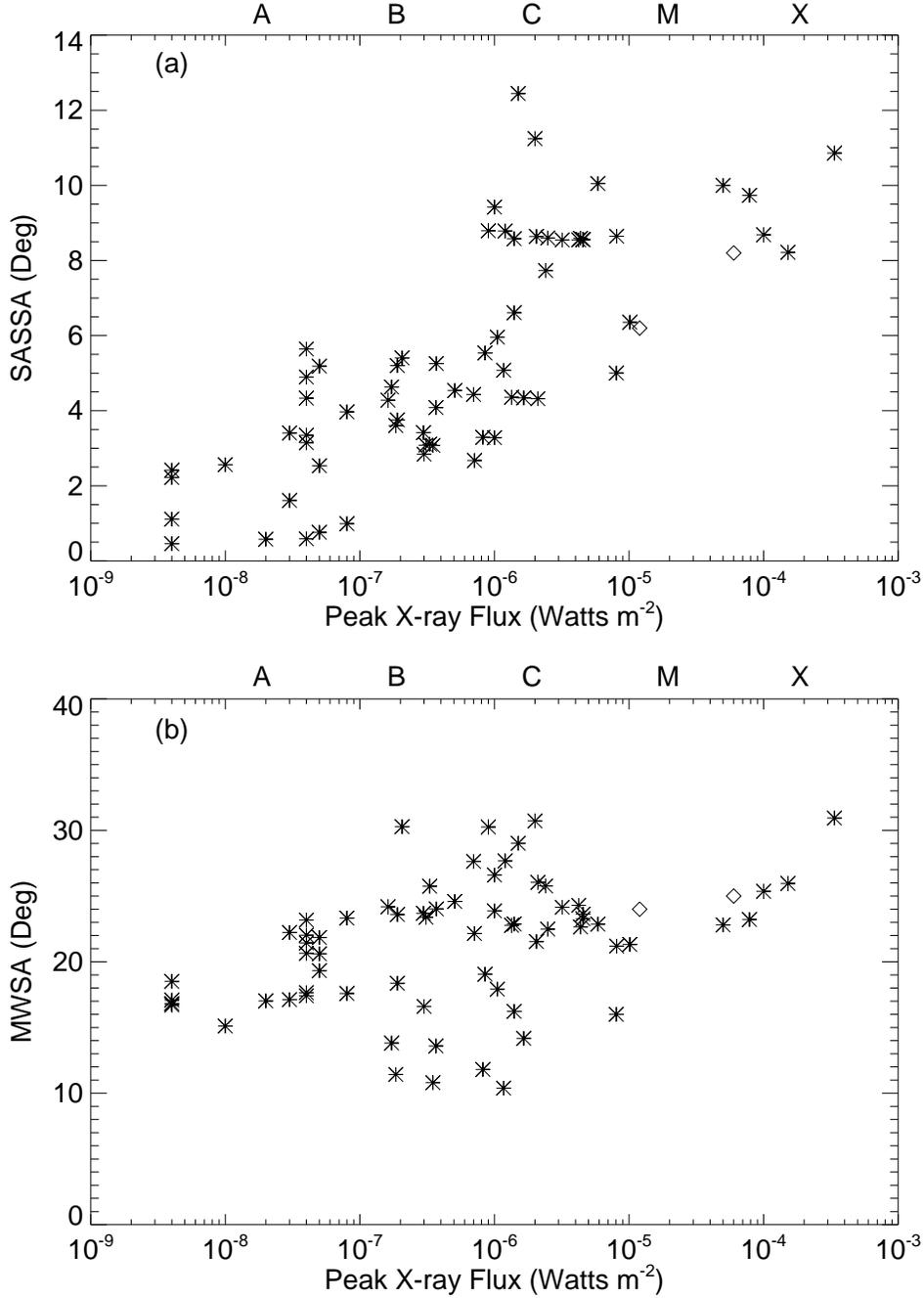}
\caption{(a): Scatter plot between SASSA and peak X-ray flux of GOES
12 satellite. Samples include all the events associated with all four
active regions i.e., NOAA ARs 10930, 10960 10961 and 10963.
The magnitude of SASSA at the time of the peak X-ray flux has been interpolated
from the available sample of SASSA as shown in Figures 3 and 4.
Also, the approximate values of the SASSA corresponding to M-class flares in two
cases have been taken from the Table 1 of \cite{tiw09b} and are shown by
diamond symbols.
(b): Same as (a) except for values of MWSA instead of SASSA.}
\end{figure}

\end{document}